\documentclass[12pt]{iopart}
\usepackage{epsf}
\begin{document}

\title{Dynamical transition for a particle in a squared Gaussian
potential}

\author{C. Touya and D.S. Dean }

\address{ Laboratoire de Physique Th\'eorique, CNRS UMR 5152, IRSAMC,
Universit\'e Paul Sabatier, 118 route de Narbonne, 31062 Toulouse
Cedex 04, France}

\begin{abstract}
We study the problem of a Brownian particle diffusing in finite 
dimensions in a potential given by $\psi= \phi^2/2$ where $\phi$ is 
Gaussian random field. Exact results for the diffusion constant in the 
high temperature phase are given in one and two dimensions and it is shown
to vanish in a power-law fashion at the dynamical transition temperature. 
Our results are 
confronted with numerical simulations where the Gaussian field 
is constructed, in a standard  way, as a sum over random Fourier modes. 
We show that when the number of 
Fourier modes is finite the low temperature diffusion constant becomes 
non-zero and has an Arrhenius form. Thus we have a simple model with a fully 
understood finite size scaling theory for the dynamical transition. In 
addition we analyse the nature of the  anomalous diffusion in the low 
temperature regime and show that the anomalous exponent agrees with that 
predicted by a trap model. 

\end{abstract}

\date{9th October 2006}
\submitto{\JPA}

\maketitle
\section{Introduction}
Systems with quenched disorder are often taken as paradigms for systems
exhibiting a structural glass transition. The basic physical idea is that
for a sufficiently complex and frustrated system without quenched disorder,
a single particle sees an effectively quenched and random potential due 
to the other particles. At a mean field level there exist models where, 
to all intents and purposes, this analogy becomes exact. 
One can have two models  one with quenched disorder and the other 
without, but highly frustrated, which exhibit the same thermodynamics in 
the high temperature phase and the same
glass transition at low temperatures \cite{fidic}. 
Even if the frustrated non-random  system possesses a crystalline 
ground state not shared by the disordered system, this fact is practically 
irrelevant as this state is dynamically never attained. 
An analogy between mean field spin glass models with one step 
replica symmetry breaking and  structural glasses has been put forward
\cite{rfo,review}, and  this theory of the glass transition has 
become known as the 
random first order  scenario. Further extensions of the  analogy 
between the dynamics of 
these mean field  spin glasses and finite dimensional glass formers have 
provided much insight into  glassy dynamics and related phenomena such as 
ageing and effective temperatures \cite{review}. 
One of the simplest models, which has been 
extensively studied, is the so called toy model. Here one takes a particle in 
a $D$ dimensional space 
with Hamiltonian
\begin{equation}
H({\bf x}) = \psi({\bf x}) + {1\over 2}\mu {\bf x}^2\ ,
\end{equation}
where $\psi$ is a quenched random potential taken from some statistical
ensemble. The most convenient choice is to take $\psi$ to be Gaussian 
of zero mean with correlation function
\begin{equation}
\langle \psi({\bf x}) \psi({\bf y})\rangle = \Delta(|{\bf x} -{\bf y}|),
\end{equation}  
so the field is statistically isotropic and invariant by translation in space. 
The statics of this model can be solved in the Gaussian variational 
approximation in finite dimensions \cite{toy1s}. This approximation
becomes exact in limit of the number of spatial dimensions $D$ 
going to infinity  if $\Delta$ scales as
\begin{equation}
\Delta(x) = D F({x^2\over D}).
\end{equation}
In this limit when the correlation function of $F$ is short range a structural
glass transition is found \cite{toy1s,toy1d}. 
First, at a temperature $T_d$ there is a  dynamical transition, which 
is independent of the precise nature of the
dynamics as long as it is  local. This dynamical transition appears 
due to the appearance of an exponentially large number of metastable states
and it can be located via  static arguments, thus explaining
its independence of the precise form of the local dynamics
\cite{toy1d}. This transition is not accompanied by
a strict thermodynamic transition.  The thermodynamic transition appears at 
a lower temperature  $T_s$ where the entropy of the system becomes very small. 
In the free case, where the harmonic confining potential is removed 
(the case $\mu = 0$), a dynamical transiting occurs at a non-zero temperature, 
but the static transition is occurs  only at $T=0$. The existence of 
the dynamical transition in the free case shows up as a transition between 
a stationary  dynamics characterised by  the time translational invariance 
of the correlation function and a form (for unbounded systems) of the 
fluctuation dissipation theorem \cite{toy1d} and a low temperature ageing 
regime exhibiting ageing in the correlation function and 
modified fluctuation dissipation relations \cite{toy1d}. In addition at high 
temperature the late time behaviour of a Brownian tracer particle in the 
force field generated by the potential is normal and is characterised by a 
non-zero  diffusion constant which vanishes at $T_d$ \cite{ddhs}. 
A little thought  convinces  one that this is not possible in finite 
dimensions and that there is thus something pathological about  the 
large $D$ limit as it is taken in 
this  model. When the correlation length of the random field is 
finite (the case where the correlation function of the field is 
infinite is another matter and anomalous diffusion is clearly possible 
in this case) one can view the
system on a coarse grained scale at the order of a few tens of this 
correlation length. We denote this length scale by $l_*$. 
The process can now be viewed as a discrete random walk
between neighbouring regions with an exponentially distributed time to jump
to a neighbouring region whose  average is  given by the Arrhenius law as 
$\tau = \tau_0\exp({\Delta E\over T})$. Here $\Delta E$ is the energy barrier
associated with moving from one region to another. When the average value of
$\tau$, $\overline \tau$ is finite the system will look like a random 
walk in the coarse grained picture and we find that
\begin{equation}
\langle {\bf X}_t^2\rangle \sim {l_*^2\over \overline{\tau}}\,t,
\end{equation}  
showing that we should have a finite diffusion constant. When $\psi$ is 
Gaussian we expect that the energy barriers $\Delta E$ are also Gaussian and
thus obtain a value of $\overline{\tau}$ which behaves as
\begin{equation}
\overline{\tau}=\tau_0\exp\left({A\over T^2}\right),\label{supa}
\end{equation} 
which is referred to as  a super-Arrhenius behaviour. 
The dynamical transition is 
however clearly relegated to zero temperature. This argument is backed up
by numerical simulations, perturbative and renormalisation
group calculations \cite{ddhs,gausdif}.  In a finite system we 
will always have finite energy barriers and they will always be overcome
by activated barrier hopping, all be it after very long but finite times.
Mean field models have diverging energy barriers and it is this divergence
which leads to the dynamical transition. Indeed the formula Eq. (\ref{supa}) 
should ultimately become Arrhenius-like for systems where the energy 
barrier are bounded and  we will discuss this point in the next section 
on exact results.

A commonly used
paradigm for glassy systems is the trap model \cite{trap1}
where the phase space is considered to be made up of a set of traps denoted by 
$i$ each of depth $\Delta E_i$. The simplest version is that where  
the traps are  on a  tree-like geometry, as is the case in the random 
energy model \cite{trap1}. The generalised random energy model has traps 
within traps and is inspired by the Parisi solution for mean field spin 
glasses where the states (corresponding to the 
bottom of traps) are organised in with an ultra-metric structure \cite{trap2}. 
The  time spent in a trap is exponentially distributed with mean time
$\tau_i =\tau_0\exp(\beta \Delta E_i)$ and the distribution of the 
$\Delta E_i$ induces the distribution of the $\tau_i$. When the 
disorder averaged value of $\tau_i= \overline{\tau}$ diverges a 
dynamical transition occurs. Forms of the trap model where the traps are 
located on a finite dimensional lattice have been extensively studied 
\cite{trapsfd,bebo}. Also a non-random  microscopic realisation, based 
on number partitioning combinatorics, of the trap model has 
been found \cite{jorge}. Intuitively we expect the trap model picture to be
applicable to the problem of a particle diffusing in a short range 
correlated random potential at sufficiently large time and length scales.
In this paper we show that for the model we study this is indeed the case. 

Going back to the problem of diffusion in a random potential,   
we consider a model where $\psi = \phi^2/2$ where $\phi$ is Gaussian.
Clearly from the arguments above we do expect to see a dynamical/glass
transition. We assume that the potential energy barriers also behave
as the energy itself, we thus assume that the statistics of the barrier 
heights behaves as $\alpha \phi^2/2$, where $\alpha$ is some positive 
constant. Therefore average time spent in a trap will behave as
\begin{equation}
\tau \sim \tau_0\exp({\beta\alpha \phi^2\over 2})\label{taud}
\end{equation}    
If we take a Gaussian field  $\phi$ with correlation function 
\begin{equation}
\langle \phi({\bf x})\phi({\bf y})\rangle = \Delta({\bf x}-{\bf y})
\end{equation}
with $\Delta(0)=1$, then $\overline{ \tau}$ diverges for $\beta\alpha >1$,
giving a dynamical transition temperature $T_g = \alpha$. Note that in 
\cite{dhs} the problem of a dipole diffusing in a Gaussian
electrostatic field $\psi$ was considered. Here the effective potential felt
by the dipole is $\psi = -{1\over 2}(\nabla \phi)^2$. The above argument
indicates that a dynamical transition, indicated by a vanishing of the 
diffusion constant, should occur in this model. 
However in \cite{dhs} no clear evidence for
a phase transition was found in numerical simulations in three dimensions. 
However standard methods for generating the  Gaussian field 
used here and in \cite{dhs} use a finite number of Fourier modes. 
Indeed we will show 
that, in the case studied here where exact results are possible, 
there is a finite size scaling in the number of modes $N$. 
This finite size scaling  smears out the dynamical transition, just as  is the 
case in the standard theory of equilibrium second order phase transitions.

The Langevin of dynamics we shall study of a particle in a potential $\psi$ 
is given by,  
\begin{equation}
\dot{\bf X}_t = {\bf \eta}(t) -\beta\nabla \psi({\bf X}_t)
\end{equation}
where $\eta(t)$ is Gaussian white noise with correlation function
\begin{equation}
\langle \eta_i(t)\eta_j(t')\rangle = 2\delta_{ij}\delta(t-t').
\end{equation}
This choice of white noise amplitude ensures that the
stationary measure is the Gibbs-Boltzmann equilibrium one. The diffusion
constant, when it exists, of  the system is defined by the late time 
behaviour of the  mean squared displacement as
\begin{equation}
\langle {\bf X}^2_t\rangle \sim 2 D \kappa_e t.
\end{equation}
Therefore using this notation the bare diffusion constant of the 
particle in the absence of the field $\psi$ will always be given by
$\kappa_e =\kappa_0 = 1$. 

We will find here that the low temperature phase is characterised by an 
anomalous sub-diffusive behaviour
\begin{equation}
\langle {\bf X}^2_t\rangle \sim  t^{2\nu},
\end{equation} 
where the exponent  associated with the anomalous diffusion $\nu < 1/2$. 
We then argue that the problem can be mapped onto an effective trap model
which allows us to predict the value of $\nu$. This prediction, in one 
dimension, is confirmed by a formal replica calculation applied to a first 
passage time problem. It is also supported by our numerical results.

\section{Exact Results}
The diffusion constant for a particle in a one dimensional random 
potential, which is statistically invariant under translation, can 
be obtained exactly  in a number of ways \cite{zwan,scat}. It is given by
\begin{equation}
\kappa_e = {1\over \langle \exp\left(\beta \psi\right)\rangle_d
\langle \exp\left(-\beta \psi\right)\rangle_d},\label{kap1d}
\end{equation}
where the subscript $d$ indicates the disorder average over the field
$\psi$ taken at any point. An interesting point about this formula
is that the resulting $\kappa_e$ is independent of the sign of $\beta$.
In the case where $\psi = \phi^2/2$ where the one-point distribution 
function for $\phi$ is Gaussian and  given by 
\begin{equation}
p(\phi) = {1\over\sqrt{2\pi}} \exp(-\phi^2/2) \label{eqpdfga}
\end{equation}
we have the result
\begin{equation}
\kappa_e = \sqrt{1-\beta^2}\label{ke1d}
\end{equation}
and so at $\beta_g = 1$ a dynamical transition thus occurs where the diffusion
constant becomes zero. This transition is really present, and below this 
transition temperature the diffusion will become anomalous. 

Note that if the energy barrier corresponding to a coarse grained  region is
given by $\alpha \phi^2\over 2$ with the distribution of $\phi$  given by
equation ({\ref{eqpdfga}),  then the induced distribution on the
trapping time $\tau$ given by equation ({\ref{taud}) is 
\begin{equation}
\rho(\tau) = {\tau_0^{1\over \alpha\beta}\over \tau^{1+{1\over \alpha\beta}}
\sqrt{\pi\alpha\beta \ln({\tau\over\tau_0})}}.\label{dist}
\end{equation}
The above trapping time distribution is almost Levy like apart from the
logarithmic term. Below the critical temperature given by $\alpha\beta_c =1$
we see that the mean value diverges, as can be seen directly. The trap model
has been extensively studied in finite dimensions in this regime. The precise
definition is as follows, the system consists of a series of traps on a
finite dimensional lattice. Associated with each site $i$ is an trap whose
residence time, the time before the next jump is made, is exponential
with quenched mean $\tau_i$. If the distribution of the $\tau_i$ are 
independent and identically distributed with distribution 
\begin{equation}
\rho(\tau) \approx A{\tau_0^\mu\over \tau^{\mu +1}}
\end{equation}
then we see that $ \overline{\tau_i}$ diverges for $\mu <1$. For 
$\mu > 1$ in one dimension we have that
\begin{equation} 
\langle {\bf X}_t^2\rangle \sim  C\ t  \ \; ;\mu >1
\end{equation}
which is normal diffusion. In the anormal phase, $\mu <1$ the mean 
squared displacement behaves as \cite{bouge}
\begin{eqnarray}
\langle {\bf X}_t^2\rangle &\sim &C\ t^{2\mu\over \mu +1} \ \; ;
D=1 \nonumber \\
&\sim &C\ t^{\mu}(\ln(t))^{1-\mu} \ \; ; D =2 \nonumber \\
&\sim & C\ t^{\mu} \ \; ; D  > 2
\label{tdiffgen}.
\end{eqnarray}

From equation (\ref{dist}) we see that up, to logarithmic corrections,
the model studied here should correspond to a trap model 
where the exponent of this anomalous diffusion is given by
\begin{equation}
\mu = \frac{1}{\alpha\beta}\label{muexp}.
\end{equation} 
The low temperature phase of the  model should thus be characterised by 
anomalous diffusion. In fact, in what follows, we shall see, both from 
analytic arguments and numerics, that the effective value of 
$\alpha$ in all the above is in fact $1$. The difficulty of trap models
with quenched disorder is that a particle may in general visit
a trap several times and thus it is not necessarily a good approximation 
to draw a new $\tau_i$ from the quenched distribution each time 
an already visited site is revisited. This approximation is called the 
annealed approximation and fails badly in one and two dimensions where the 
random walk is recurrent \cite{bouge}. The model corresponding to the 
annealed approximation is the annealed model where each time a particle
visits a site the trapping time is redrawn from the distribution of 
waiting times (which is site independent). 
Above the critical value
$\mu_c=1$ the annealed model in one dimension diffuses as  
\begin{equation} 
\langle {\bf X}_t^2\rangle \sim  C\ t + O(\sqrt{t})\ \; ;\mu >2.
\label{anne1}
\end{equation}
However, still above $\mu_c =1$ but with  $1<\mu< 2$ we find 
\begin{equation}
\langle {\bf X}_t^2\rangle \sim  C\ t + O(t^{1\over \mu}) \; ;1< \mu <2.
\label{anne2}
\end{equation}
This change in the exponent of the sub-leading temporal behaviour 
stems from the fact that the variance of the time spent in 
a trap  ${\overline  \tau_i^2} -{\overline  \tau_i}^2 $ diverges,
while the mean value remains finite. Thus we see that there are two regimes
of normal diffusion, a high temperature one corresponding to $\mu >2$ 
as given by equation ({\ref{anne1}) and a low temperature normal phase
with  $1<\mu<2$ and where the corrections to pure diffusion change from 
$O(\sqrt{t})$ to $O(t^{\frac{1}{\mu}})$. The key point is the divergence of the
variance of the trapping time at a given site and we thus expect a similar
transition to occur in the quenched version of the model. 
For $\mu <1$ the annealed model gives
\begin{equation}
\langle {\bf X}_t^2\rangle \sim C t^{\mu}
\end{equation}
in all dimensions. Thus we see that, in low dimensions,  
the exponent associated with the anomalous diffusion in the 
annealed model is not the same as that
associated with the quenched model. This is precisely due to the recurrence of
random walks in two and less dimensions.

For our later 
comparison with numerical simulations we will need to compute the 
correction to normal diffusion in the high temperature limit. As random
walks are recurrent in one dimension transient effects are very important 
and unless one can predict them the extraction of a diffusion constant from
a numerical simulation becomes extremely difficult. 

In the high temperature phase we write
\begin{equation}
\langle {\bf X}_t^2\rangle \sim 2\kappa_e t + O(t^\theta).
\end{equation}
Now if  the corresponding trap model is characterised by $\alpha =1$ 
then we have $\mu = 1/\beta$. Adapting the scaling argument found in  
\cite{bouge} to the finite time corrections we predict that  
\begin{eqnarray}
\theta &=& 3/4\  {\rm for} \ \mu > 2 \nonumber \\
        &=& {1\over 2} +{1\over 2\mu}  = {\beta +1\over 2 }\ {\rm for}\  
1\leq \mu\leq 2,
\label{eqtheta}
\end{eqnarray} 
The above results for the exponent $\theta$ turn out to be in good
agreement with our numerical results. This supports both our scaling arguments
and the identification of our model with a trap model characterised by
the exponent $\mu = 1/\beta$ (equivalently $\alpha =1$).  
The prediction of equation (\ref{eqtheta})
shows that the finite time corrections to diffusion are always important, and
as one approaches the transition at $\beta =1$ they become of the same 
order as the normal diffusion term.

Returning to the Langevin problem one may be tempted to argue that
the transition seen in one dimension could be rather pathological in that the 
system is obliged to overcome all energy barriers in one dimension. 
Unfortunately no general exact results exist in higher dimensions, however 
in two dimensions if the field $\psi$ is statistically the same as $-\psi$
a duality argument can be used \cite{ddhd} (in a rather indirect way) 
to show that 
\begin{equation}
\kappa_e = {1\over
\langle \exp\left(-\beta \psi\right)\rangle_d}.\label{kap2d}
\end{equation}
This clearly allows us to compute $\kappa_e$  in two dimensions in the case 
where
$\psi$ Gaussian. We can however get an exact result for another potential 
by judiciously choosing  a field of the form 
\begin{equation}
\psi = {\phi^2\over 2}- {\phi'^2\over 2}, \label{2dex}
\end{equation}
where $\phi$ and $\phi'$ are independent Gaussian fields with the same 
statistics. Clearly $\psi\sim-\psi$ in the statistical sense and we 
thus obtain the  exact result
\begin{equation}
\kappa_e = \sqrt{1-\beta^2},
\end{equation}
showing that the same dynamical transition occurs in two dimensions for this
particular choice of $\psi$.

If one wants to simulate the Langevin process considered here on a computer,
without resorting to a lattice model, one may use a standard method to 
generate a Gaussian random field \cite{modes}, where one takes 
\begin{equation}
\phi = \sqrt{{2\over N}}\sum_{n=1}^N \cos({\bf k}_n\cdot {\bf x} + \epsilon),
\label{phifs}
\end{equation}
 where $N$ is the number of modes chosen. When, as we  assume here, 
$\Delta(0) = 1$ one takes each wave vector ${\bf k}_n$ 
independently from the distribution
\begin{equation}
P({\bf k}) = 
{1\over (2\pi)^D}\int d{\bf x}\,\exp(i{\bf k}\cdot{\bf x})\Delta({\bf x}).
\end{equation}
For instance here we will carry out simulations with
\begin{equation}
\Delta({\bf x}) =\exp(-{{\bf x}^2\over 2})
\end{equation}
and so each component of a vector ${\bf k}_n$ is a Gaussian of mean zero
and variance one. In the limit of  large $N$ the field so generated becomes
Gaussian but it is interesting to see what the effect of a finite number of
modes is from both a theoretical and practical (for comparison with numerical
simulations) point of view.  We start with the simple Gaussian case where 
we have 
\begin{equation}
\langle \exp\left(\beta \phi\right)\rangle_d  =
\left[ {1\over 2\pi}\int_0^{2\pi} d\epsilon\,
\exp\left(\beta \sqrt{2\over N}\cos(\epsilon)\right)\right]^N
= I_0\left( \beta \sqrt{2\over N}\right)^N\ ,
\end{equation}
where $I_0$ is a modified Bessel function \cite{abram}. The function 
$I_0$ has the small $z$ behaviour given by \cite{abram} 
\begin{equation}
I_0(z) = 1 + {{1\over 4}z^2\over (1!)^2} + {({1\over 4}z^2)^2\over (2!)^2}
+ O(z^6)\ ,
\label{I0sz}
\end{equation}
and the large $z$ asymptotic behaviour \cite{abram}
\begin{equation}
I_0(z) =  {\exp(z)\over \sqrt{2\pi z}}\left[ 1 + O({1\over z})\right].
\label{I0lz}
\end{equation}
From equation ({\ref{I0sz}) we thus find that in the limit 
$N\to\infty$ but keeping $\beta$  finite  we find
\begin{equation}
\kappa_e = \exp(-{\beta^2}) = \exp(-{1\over T^2})
\end{equation}
in one dimension. We thus obtain the super-Arrhenius law associated with 
as expected. However at much lower temperatures 
such that  $\beta \gg N^2$ we use the asymptotic form equation (\ref{I0lz}) to obtain
\begin{equation}
\kappa_e = 
\left( {8 \pi^2 \beta^2\over N}\right)^{1\over 2} 
\exp\left(- 2\beta\sqrt{2 N}\right) \sim \exp(- {2\sqrt{2 N}\over T})
\label{alaw1d}
\end{equation}
and thus we recover the Arrhenius law at extremely low temperatures. Note that
the term $\Delta E$ in equation (\ref{alaw1d}) is simply the maximal 
difference in energy possible for the field $\phi$ expressed as a finite
Fourier series as in equation (\ref{phifs}).  

As stated earlier when $\psi = \phi$ we have the statistical equivalence 
necessary to use the result equation (\ref{kap2d}) and we find the 
corresponding two dimensional results
\begin{equation}
\kappa_e = \exp(-{\beta^2\over 2}) = \exp(-{1\over  2T^2})
\end{equation}
for $N\to \infty$ and $\beta$ finite , and 
\begin{equation}
\kappa_e = \left( {8 \pi^2 \beta^2\over N}\right)^{1\over 4} 
\exp\left(- \beta\sqrt{2 N}\right) \sim \exp(- {\sqrt{2 N}\over T}),
\label{alaw2d}
\end{equation}
for $\beta \gg N^2$. We see again that the Arrhenius law is recovered
in this limit, but in contrast with the one dimensional case, the term 
$\Delta E$ in equation (\ref{alaw2d}) is only half of the maximal energy
difference between two points. This makes sense as in two dimensions one
can {\em go around} this maximal energy barrier.

The case of a squared Gaussian potential is treated similarly using
\begin{eqnarray}
\langle \exp(\beta \psi)\rangle_d &=& 
\int_{-\infty}^{\infty} {dz\over \sqrt{2\pi}}
\exp(-{z^2\over 2})\langle\exp(z\sqrt{\beta}\phi)\rangle_d \nonumber \\
&=& \int_{-\infty}^{\infty} {dz\over \sqrt{2\pi}}
\exp(-{z^2\over 2})I_0\left(z\sqrt{2\beta \over N}\right)^N \nonumber \\
&=& \sqrt{{N\over 2\pi}}\int_{-\infty}^{\infty} du \exp\left(N f(u)\right)
\label{sp1}
\end{eqnarray}
where
\begin{equation}
f(u) = -{u^2\over 2} +\ln\left[I_0 \left(u\sqrt{2\beta}\right)\right].
\end{equation}
The integral in equation ({\ref{sp1}) can now be treated in the saddle point
approximation. From the small $z$ expansion of $I_0$ of equation (\ref{I0sz})
we see that, about $u=0$,  $f(u)$ takes the form
\begin{equation}
f(u) = - (1-\beta){u^2\over 2} - \beta^2 {u^4\over 16} +O(u^6).
\end{equation}
We therefore see that the dynamical transition occurring at $\beta =1$ is 
mathematically equivalent to a mean field ferromagnetic transition !
Note that if $\beta$ is negative then no transition occurs so the 
term $\langle \exp(-\beta\psi)\rangle$ behaves analytically as $N\to \infty$.
In the high temperature phase $\beta <1$ therefore we have
\begin{equation}
\langle \exp(\beta \psi)\rangle_d = {1\over \sqrt{1-\beta}}\left[
1 - {3\over N(1-\beta)^2} + O({1\over N^2})\right]\ ,
\end{equation}
which gives
\begin{equation}
\kappa_e = \sqrt{1-\beta^2}\left[1 + {3\beta^2\over 8 N}{\beta^2 +1\over (1-\beta^2)^2}+ O({1\over N^2})\right].
\end{equation}
When $\beta >1$ the saddle point is no longer at $u=0$ and the function 
$f$ has a maximum value greater than zero at the points $\pm u_c$ where
the maximum is attained. Here we  find that
\begin{equation}
\kappa_e = {1\over 2}\sqrt{(1+\beta)|f''(u_c)|}\exp\left(-N f(u_c)\right).
\end{equation}
For $\beta\sim 1$ we find that $u_c = 2\sqrt{\beta-1}/\beta$ and consequently
\begin{equation}
\kappa_e = {1\over 2}\sqrt{2(\beta^2-1)}
\exp\left(-N {(\beta-1)^2\over \beta^2}\right).
\end{equation} 
At low temperatures we can use the asymptotic behaviour in equation 
({\ref{I0lz})
to obtain $u_c\approx \sqrt{2\beta}$ and thus
\begin{equation}
\kappa_e = {1\over 2}\sqrt{(1+\beta)}\exp\left(-\beta N\right)\sim \exp
\left(-{N\over T}\right).
\end{equation}
Again note that that the last part of the above equation indicates an 
Arrhenius law corresponding to the maximum difference in energy between two
points, the lowest value of $\psi$ being zero and the maximum being $N$.   
In  two dimensions for the symmetric potential of equation
(\ref{2dex}) we find exactly the same results as for the one dimensional 
potential. 

To give the reader a feel for what happens when the  number of
modes is finite we have plotted the $N=\infty$ result, equation 
(\ref{ke1d}), against the  corresponding results for $N=64$ and $N=128$, 
often considered to be sufficient for simulation purposes, 
which are evaluated numerically using the exact equation (\ref{sp1}).
We see that even for $N=128$ that the value of $\kappa_e(1)$ is still of
order 1 ! 

Given the above, it is interesting to examine the case
of a diffusing dipole as studied in \cite{dhs}. 
In the case of positive temperature the potential  
$\psi = -{1\over 2}(\nabla \phi)^2$ has localised minima of varying depths 
which can be regarded as traps. Therefore in this case one would think that 
the trap model picture would be a  reasonably good approximation. 
If we take the equation ({\ref{taud}) with
$\alpha=1$ and apply it to this case (authors of \cite{dhs} used 
$\Delta({\bf x}) = \exp(-{\bf x^2}/2) $) one finds that the transition
occurs at $\beta =1$. Interestingly the numerical results of \cite{dhs}
indicate a cross over from a concave behaviour to a convex behaviour of
$\kappa_e$ in this region, rather reminiscent of the behaviour of 
the finite $N$ values in figure (\ref{figkn}) near the dynamical transition.
It is also interesting to note that the third and fourth order perturbation
theory results of \cite{dhs} give values of $\kappa_e$ which vanish close
to $\beta =1$ (extremely close in the case of the third order result).
Having said this we have no rigorous proof that there should be a 
transition in the model studied in \cite{dhs}. However, for the reasons 
presented here, a finite size scaling analysis could resolve this issue. 
In \cite{dhs} there is no evidence for a transition at negative temperatures. 

\begin{figure}
\begin{center}
\epsfxsize=0.7\hsize\epsfbox{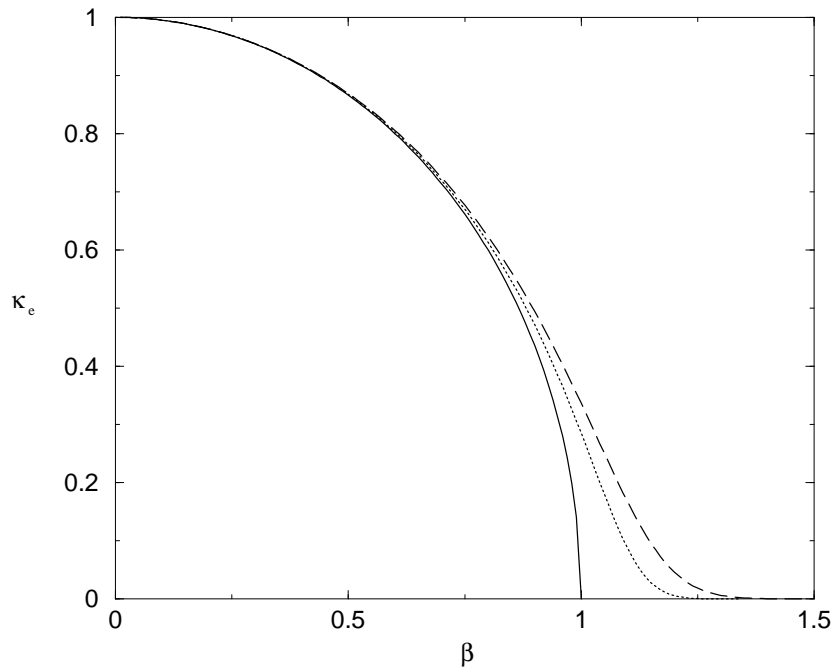}
\end{center}
\caption{Exact value of the diffusion constant $\kappa_e$ in one
dimension for a squared potential. The solid line
shows the exact Gaussian result ($N=\infty$ modes) and the dotted 
and dashed lines are for $N=128$ and $N=64$ modes respectively. Note that
at the Gaussian transition temperature given by $\beta=1$ the corresponding
$\kappa_e$ for $N=128$ modes is $\kappa_e(1)\approx 0.3$ }
\label{figkn}
\end{figure}

To summarise, our exact results have shown that, in the squared
Gaussian cases studied here, the high temperature regime is
characterised by a relaxation time $\tau$ (using the fact that the
correlation length scale is O(1)) which is finite but diverges as
$T\to T_g$ as
\begin{equation}
\tau \sim {1\over (T-T_g)^{{1\over 2}}}.
\end{equation}
However when the potential is constructed from a finite number of modes 
we find that for $T<T_g$. 
\begin{equation}
\tau \sim \exp({N\Delta \epsilon(T)\over T})
\end{equation}
where $N\Delta \epsilon (T)$ can be interpreted as a temperature dependent
energy barrier. In one dimension this barrier height tends to the maximal 
difference in potential possible between two points as the temperature 
approaches zero and in two dimensions it tends to half the maximal 
difference.    
  
In one dimension one can obtain information about transport behaviour by
analysing first passage time. This method can be used to 
calculate the diffusion constant in a one dimensional potential \cite{zwan}
and was used in \cite{ledou} to determine the exponents of the anomalous
diffusion in a Gaussian potential with long range (logarithmic) correlations. 
To simplify the calculation one considers diffusion on the half line 
with reflecting
boundary conditions at the origin $0$. We denote by $T(L)$ the average
(in the thermal sense of averaging over realisations of the white noise)
time at which the tracer particle starting at the origin first reaches the 
point $L$. This average first passage time is given by (see \cite{ledou}
and references within) the following double integral which depends on the
realisation of the driving field $\psi$
\begin{equation}
T(L) = \int_0^L dy dx \;\theta(y-x) \exp\left(\beta\psi(y) -\beta\psi(x)\right).\label{eqfpt}
\end{equation}
In general we can compute the moments of $T$ by replicating the double 
integrals to yield upon disorder averaging
\begin{equation}
\overline{ T(L)^n} = \int_0^L \prod_a dy_a dx_a \theta(y_a-x_a)
\det\left(I+\beta A \Lambda\right)^{-{1\over 2}},\label{reps}
\end{equation}
where the matrices $A$ and $\Lambda$ are symmetric $2n\times 2n$ matrices with
\begin{equation}
A_{ab} = \Delta(y_a-y_b) 
\end{equation}
and 
\begin{eqnarray}
\Lambda_{ab} &=& -\delta_{ab} \ \ \  {\rm for}\  1\leq a,b\leq n    \nonumber \\
&=& 0\  \ \ {\rm for} \ 1\leq a\leq n, \ n+1\leq b\leq 2n\nonumber \\
&=&  \delta_{ab}\ \ \ {\rm for} \  n+1\leq a,b \leq 2n,
\end{eqnarray}
where we have used the notation $x_a = y_{n+a}$. For the first moment $n=1$ we
find, where it is finite, that
\begin{equation}
\overline{ T(L)} =  \int_0^L \prod dy dx\;  \theta(y-x)
{1\over \left(1 -\beta^2 +\beta^2  \Delta^2(y-x)\right)^{1\over 2}}\label{eqt1}
\end{equation}
For $\beta <1$ we may write
\begin{equation}
\overline{ T(L)} = {1\over 2 L^2 \sqrt{1-\beta^2}} +
\int_0^L  dy dx \; \theta(y-x) 
\left({1\over \left(1+ {\beta^2\over 1-\beta^2} 
\Delta^2(y-x)\right)^{1\over 2}} -1\right),
\end{equation}
which becomes for large $L$,
\begin{equation}
\overline{ T(L)} = {L^2\over 2  \sqrt{1-\beta^2}} - L C\label{tpp}
\end{equation}
where
\begin{equation}
C = \int_0^\infty dz \;\left[1-{1\over 
\left(1+{\beta^2\over 1-\beta^2} \Delta^2(z)\right)^{1\over 2}}\right]
\end{equation}
when it is finite. The leading order in $L$ of the right hand side above 
yields the correct high temperature diffusion constant.

Inspection of the double integral in equation ({\ref{eqt1}) shows that, 
assuming that $\Delta(x)$ is monotonically decreasing, the value of
${\overline T(L)}$ is finite as long as $\Delta(L)^2 > (\beta^2-1)/\beta^2$.
This is clearly always the case in the high temperature phase $\beta <1$. 
In the low temperature phase there is now a length scale $L_c$, which is 
temperature dependent: for $L<L_c$ $\overline{ T(L)}$ is finite, and 
for $L>L_c$ it is divergent. For a monotone $\Delta$ this means that $L_c$
decreases with temperature and we may tentatively relate this with the 
idea of the system progressively freezing on smaller and smaller length
scales as the temperature is decreased. This image of the system becoming 
frozen on smaller and smaller length scales is one which has been used
extensively to interpret experiments on spin glasses \cite{length}. 

We will now present an argument which will give a prediction for the
exponent of anomalous diffusion in the low temperature phase. The
argument is similar in spirit to that of \cite{ledou} for long range
Gaussian potentials in one dimension, although it is difficult to make
it as rigorous as in the long range case. First notice that as one
approaches the transition temperature the terms causing the divergence
in equation (\ref{eqt1}) are those where $x$ and $y$ are far away from
each other and hence uncorrelated. Secondly note that in the equation
(\ref{eqfpt}) we expect (in the case of $\beta >0$) that the first
passage time is dominated by the maximal value of $\phi$. Let us
denote this value by $y_*$. In the replicated averaged formula
equation ({\ref{reps}) we thus expect the most divergent term to occur
at $y_a = y_*$ for all $a$. Also we expect that their are many values 
of $x$ which contribute to the most divergent term in equation ({\ref{reps})
and these are points where $\phi(x)=0$. In general theses points will be well
separated and far from $y_*$. Thus over most of the replicated interval
$|x_a-x_b| \gg1 $ and $|x_a - y_*| \gg1$. This means that one can make the 
approximation in the integral that $\Delta(x_a -x_b) \approx 0$ and 
$\Delta(x_a -y_*) \approx 0$ and thus the replicated integral in
equation (\ref{reps}) can be approximated as
\begin{eqnarray}
\overline{ T(L)^n} &\approx &\int_0^L dy_*\int_0^{y_*} dx_a
\det\left(I+\beta B \Lambda\right)^{-{1\over 2}} \\ \nonumber
&\approx & \det\left(I+\beta B \Lambda\right)^{-{1\over 2}}
\ {L^{n+1}\over n+1}
\end{eqnarray}
where
\begin{eqnarray}
B_{ab} &=& -1 \ \ \  {\rm for}\  1\leq a,b\leq n    \nonumber \\
&=& 0\  \ \ {\rm for} \ 1\leq a\leq n, \ n+1\leq b\leq 2n\nonumber \\
&=&  \delta_{ab}\ \ \ {\rm for} \  n+1\leq a,b \leq 2n,
\end{eqnarray}
The eigenvalues of $B$ are easily calculated, there being $1$ equal to
$-n\beta$, $n-1$ equal to $0$ and $n$ equal to $1$.

This now gives
\begin{equation}
\overline{ T(L)^n} 
\approx {1\over (1+n)(1-n\beta)^{1\over 2}(1+\beta)^{n\over2}} L^{n+1}
\label{tnap}
\end{equation}

Now if we choose the exponent $n$ to be very close to ${1\over\beta}$ we
find that the equation (\ref{tnap}) is indeed diverging and should 
be the dominant contribution. This  yields
\begin{equation}
 \overline{ T(L)^{1\over \beta}} \sim L^{1+{1\over \beta}}
\end{equation}
and thus dimensionally we have
\begin{equation}
\langle X_t^2\rangle \sim t^{2\over 1+\beta}
\end{equation}
in agreement with the arguments in \cite{bebo} and references therein.

In the presence of an external field the 
potential becomes ${1\over 2} \psi^2 - hx$. The same line of reasoning now 
gives for large $L$
\begin{eqnarray}
\overline{ T(L)^n} &\approx &\int_0^L dy_*\int_0^{y_*} dx_a 
\exp\left(-nh y_* + \sum_a hx_a\right)
\det\left(I+\beta B \Lambda\right)^{-{1\over 2}} \\ \nonumber
&\approx & \det\left(I+\beta B \Lambda\right)^{-{1\over 2}} 
{L\over (\beta h)^n}.
\end{eqnarray}
Again we set $n$ very close to ${1\over \beta}$
and dimensionally find the form
\begin{equation} 
\langle X_t\rangle \sim h^{1\over \beta}t^{1\over \beta}
\end{equation}
again in agreement with the results of \cite{bebo}, not only for the temporal
exponent but also for the exponent associated with $h$. The above calculation
also show that the value of $\alpha$ associated with the effective trap model
defined by equation ({\ref{taud}) for this problem is indeed $\alpha=1$. 
\section{Numerical Simulations} 
In this section we will verify the analytical results of the previous sections
via two distinct types of numerical simulations. We begin with a  direct 
simulation of the Langevin process using a second order Runga-Kutta 
integration of the Langevin equation as developed in references \cite{sdenum}
and generating the random field $\phi$ via equation ({\ref{phifs}) as
proposed originally in \cite{modes}. This simulation technique allows access to
the asymptotic regime, where the relevant transport coefficients can be 
evaluated, at high temperatures in the regime of normal diffusion and 
at temperatures below  $T_g$, but not too low. Clearly in order
to attain the asymptotic regime at low temperatures we need to diffuse
over a sufficiently large distance. However as the diffusion becomes 
progressively slower the times needed to attain this regime become  
prohibitive from the computational point of view. We expect that
the low temperature phase should be described by a one dimensional trap model 
but, as discussed above, it is not completely obvious what parameters 
we should take for this trap model.
We have therefore constructed an effective  trap model directly 
from a realisation of the  random field as generated by equation 
(\ref{phifs}). The procedure used
is the following. We take a realisation of the field $\psi$ over a suitably 
large interval. Each minima $\alpha$ of the field is associated with a 
point $\alpha$ on the line at its actual position  denoted by $x_\alpha$.
We then calculate the energy barrier to move to the nearest minima to the 
left $\alpha-1$ and the right $\alpha+1$. These two energy barriers are
denoted by $\Delta E_{\alpha}^{(L)}$ and $\Delta E_{\alpha}^{(R)}$
respectively. The particle then is taken to jump to the left/right at
an exponentially distributed random time $T_{L/R}$ with average value given
by the Arrhenius law $\tau_{L/R} = \exp\left( \beta \Delta E_{\alpha}^{(L/R)}
\right)$ (the over all scaling of time is unimportant to determine the 
exponent associated with the anomalous diffusion). From the site $\alpha$
$T_{L/R}$  are generated numerically and the particle hops to the site 
corresponding to the shorter time. The time of the simulation is then
increased by this shorter time. In this way we can assure that the system 
has diffused sufficiently far and the total simulation time is independent
of the physical time, allowing us to attain the asymptotic regime.     

Our numerical integration  of the stochastic differential in the 
high temperature normal phase yields the following results. 
Shown in figure  ({\ref{fitk}) is the value of the diffusion constant
$\kappa_e$ determined from a fit $\langle X^2_t\rangle = 
2\kappa_e t + Bt^\theta$, where $\theta$ is 
the exponent predicted by equation (\ref{eqtheta}). The numerical result is generated by averaging over $10000$ 
particles in each realisation of the  random field and averaging 
over $100$ realisations generated using $128$ modes. 
The total time of the simulation consisted  of $5000$ integration steps 
using $\Delta t = 0.5$ in the second order  stochastic Runga-Kutta. 
Again we see that the above fit is in  good  agreement with the theoretical 
result for $N=128$ (dashed line) modes. Also shown is the 
$N\to\infty$ result (solid line) 
\begin{figure}
\begin{center}
\epsfxsize=0.7\hsize\epsfbox{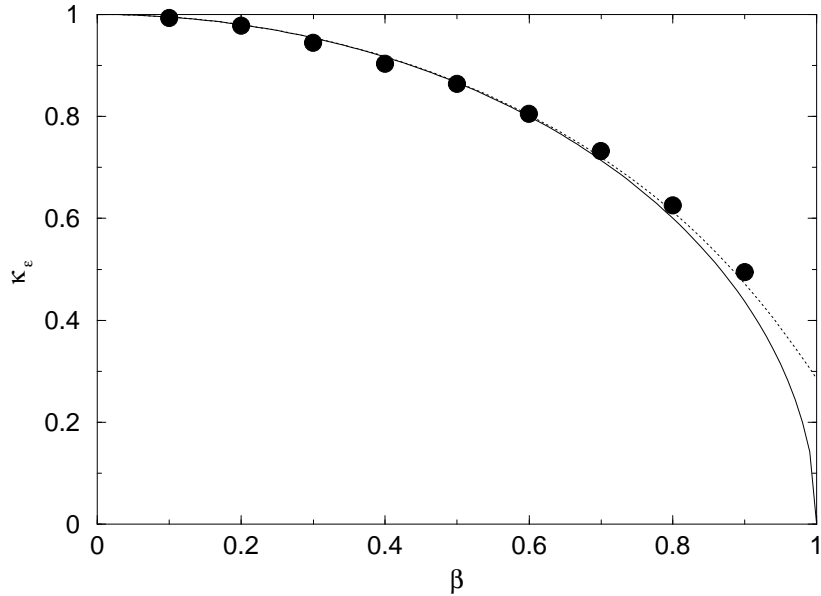}
\end{center}
\caption{Fit of simulation results (circles), 
assuming an exponent $\theta$ given by  equation (\ref{eqtheta}), 
of the diffusion constant $\kappa_e$ in one dimension using
$N=128$ modes. The dashed line shows the analytic result 
for $N=128$ modes and the solid line that in the limit $N\to \infty$.}
\label{fitk}
\end{figure}

In figure (\ref{fittheta}) we show the fitted value of the exponent
$\theta$ assuming the exact result (for $N=128$ modes) for $\kappa_e$. We see
that the result is in good accord with the scaling prediction of equation 
(\ref{eqtheta}) for $\theta$. 

\begin{figure}
\begin{center}
\epsfxsize=0.7\hsize\epsfbox{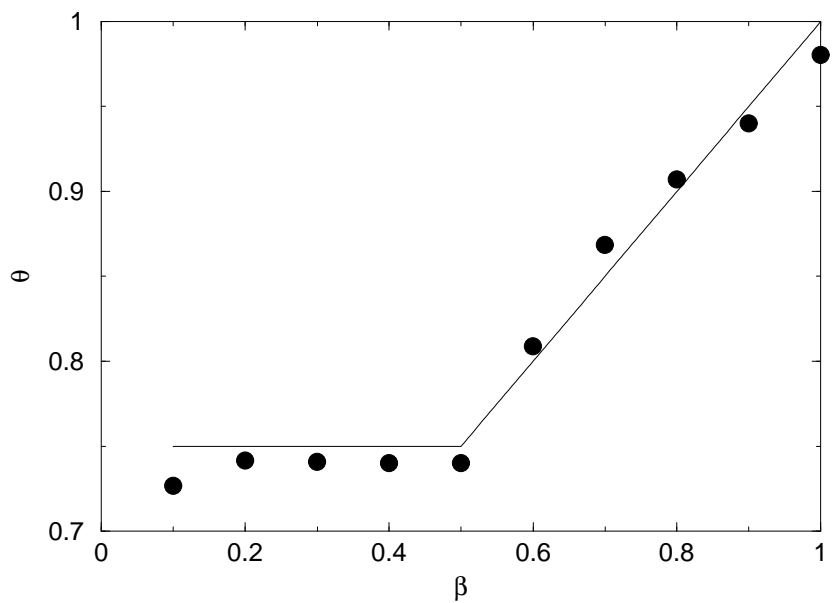}
\end{center}
\caption{Fit of simulation results (circles), 
of the exponent  $\theta$,  using the analytic result for $\kappa_e$ in one dimension with  $N=128$ modes. The solid line shows the scaling prediction equation
(\ref{eqtheta}) for $\theta$.}
\label{fittheta}
\end{figure}

For temperatures not too far below $T_g$ we may use a direct numerical 
integration to estimate the exponent $\nu$. The results of these simulations
are shown in figure (\ref{fitnu}) (empty squares). We see that the 
numerically measured value of the exponent is close to the predicted one
up to  $\beta\sim 1.5$ but after it departs from the predicted value. We
believe that this because we are not  carrying out the simulation for 
sufficiently long times. Indeed when we simulate the trap model, where
simulation time is no longer a problem, we see from  figure (\ref{fitnu})
(filled circles) that the agreement with the theoretical result is much 
improved for the  larger values of $\beta$.

\begin{figure}
\begin{center}
\epsfxsize=0.7\hsize\epsfbox{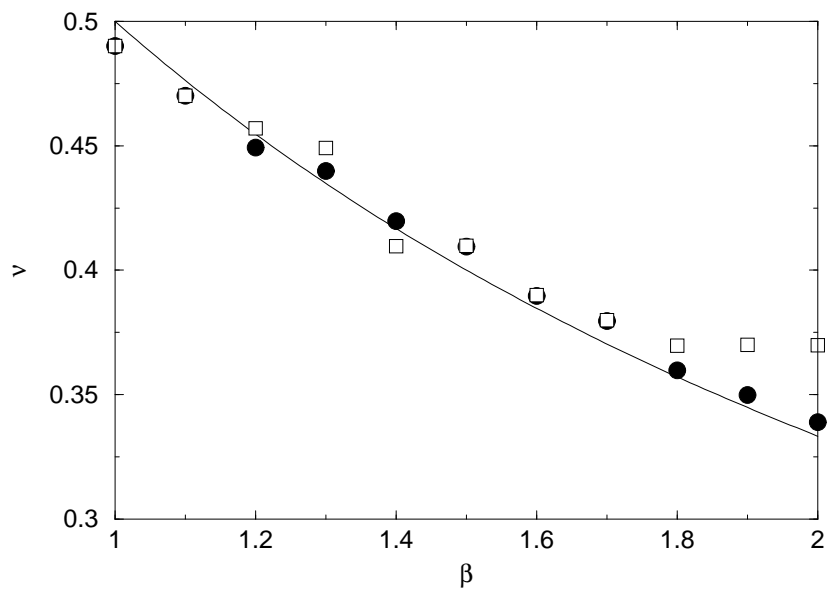}
\end{center}
\caption{Fit of exponent of anomalous diffusion $\nu$ 
obtained from the one dimensional effective trap model (circles) and 
direct integration of the stochastic differential equation (squares) shown 
against 
the prediction $\nu = 1/(1+\beta)$ (solid line).
}
\label{fitnu}
\end{figure}
\section{Conclusions}
We have studied the dynamics of a particle diffusing in an potential which is
given by the square of a Gaussian potential whose statistics are translationally invariant in space and whose correlation function  
is short ranged. In contrast to the Gaussian case, there really is a 
dynamical transition for this model. The transition manifests itself as a
crossover between a high temperature diffusive regime and a low temperature 
regime where the particle diffusion is anomalous and more specifically 
is sub-diffusive. We showed how the diffusion constant could be computed in
one dimension and in a special case in two dimensions. Interestingly when 
the Gaussian field is constructed using a finite Fourier series we see that,
instead of vanishing at the dynamical transition temperature, the diffusion 
constant obeys an Arrhenius form dependent on the maximal
energy barrier present in the system, Explicitly we have shown
\begin{equation}
\kappa_e \sim C \exp(- {AN\over T}),
\end{equation}
at low temperatures. 
This result is physically intuitive and 
our calculation allows for a complete understanding of the finite size 
scaling underlying this dynamical transition. We also showed that these
finite scaling effects can be important for the number of modes typically 
used in numerical simulations.
In addition, in one dimension
we show that the low temperature phase can be described in terms of a trap 
model where the energy barriers are assumed to have the same statistics as
the energy function itself. The resulting sub-diffusive behaviour for the
low temperature regime in one dimension is
\begin{equation}
\langle X^2_t\rangle \sim t^{2\over 1+\beta}.
\end{equation}    
This scaling, form and that for the mean displacement in the presence
of a uniform drift was also obtained via a, non-rigorous, replica
based computation of the moments of the mean first passage
time. Finally in the case of a finite number of modes it is a natural
question ask which time scale must one go to in order to see the
finite size effect diffusive behaviour in the low temperature regime
as opposed to anomalous diffusion. A simple way to estimate this cross
over time $t_c$ is to equate the mean squared displacements in the
anomalous and diffusive cases, i.e.
\begin{equation}
t_c^{2\over 1+\beta} \sim      \exp(- {AN\over T})\ t_c,
\end{equation}        
which yields at low temperatures
\begin{equation}
t_c\sim \exp\left({AN\over T}\right).
\end{equation}

\noindent{\bf Acknowledgments}: DSD would like to thank the Isaac Newton
Institute Cambridge where part of this work was carried out during the 
programme  Principles of the  Dynamics of Non-Equilibrium Systems. We would 
also like to thank J.-P. Bouchaud for useful conversations and pointing 
out the interest of this model.

\end{document}